\documentclass[preprint,aps]{revtex4}
 \usepackage{amsfonts}
   \usepackage{amssymb}
\begin{document}
\title{
Theory of pseudogaps in charge density waves in
 application to photo electron or tunneling spectroscopy.}
\author{S.I. Matveenko$^{2,1}$ and S. Brazovskii$^{1,2}$ }
\address{$^{1}$Laboratoire de Physique Th\'{e}orique et des Mod\`{e}le Statistiques,\\
CNRS, B\^{a}t.100, Universit\'{e} Paris-Sud, 91405 Orsay, France. \\
$^{2}$L.D. Landau Institute for Theoretical Physics,\\
Kosygina Str. 2, 119334, Moscow, Russia.}
\date{21 May 2003}
\bigskip
\published[ Written   for
Proceedings of ECRYS-2003 \cite{ECRYS02}; Published  as \cite{matv-ecrys}]
\bigskip
\bigskip
\bigskip
\begin{abstract}
For a one-dimensional electron-phonon system we consider the
photon absorption involving electronic excitations within the
pseudogap energy range. In the framework of the adiabatic
approximation for the electron - phonon interactions these
processes are described by nonlinear configurations of an
instanton type. We calculate the subgap absorption as it can be
observed by means of photo electron or tunneling spectroscopies.
In details we consider systems with gapless modes: 1D
semiconductors with acoustic phonons and incommensurate charge
density waves. We find that below the free particle edge the
pseudogap starts with the exponential decrease of transition rates
changing to a power law deeply within the pseudogap, near the
absolute edge.
\end{abstract}

\bigskip
\maketitle

\section{Pseudogap, solitons, instantons.}

This  article  is  devoted  to  theory  of  pseudogaps (PGs) in
applications to Photo  Electron  Spectrography  (PES).  We  shall
study  influence  of quantum lattice  fluctuations  upon
electronic  transitions  in  the subgap region for one-dimensional
($1D$)   systems   with   gapless   phonons. We  will  show  that
sound branches of phonon spectra change drastically transition
rates  making  them  much  more pronounced   deeply   within   the
PG. We   shall   consider  generic  $1D$ semiconductors  with
acoustic  {\it  e-ph}  coupling and Incommensurate Charge Density
Waves  (ICDWs)  which  possess  the  gapless  collective  phase
mode.

Details and illustrations can be found in  \cite{sound-jetp}.
 Low   symmetry systems  with  gapful  spectra  have  been
addressed  by  the authors earlier \cite{matv-01} and we refer to
this article for a more comprehensive review and references.

The PG concept introduced in \cite{lra}  refers today to various
systems where a gap in their bare electronic spectra is partly
filled showing subgap tails. Even for pure systems and at zero
temperature there may be a rather smeared edge $E_{g}^{0}$ while
the spectrum extends deeply inwards the gap till some absolute
edge $E_{g}<E_{g}^{0}$ which may be even zero (no true gap at
all). A most general reason is that stationary excitations
(eigenstates of the total {\it e-ph} system) are the self-trapped
states, {\em polarons} or {\em solitons} \cite{braz-84} which
energies, $W_{p}$ or $W_{s}$, are below the ones of free electrons
thus forming $E_{g}=W_{p},W_{s}$. Nonstationary states filling the
PG range $E_{g}^{0}>E>E_{g}$ can be observed {\em only via
instantaneous measurements} like optics, PES or tunneling.
Particularly near $E_{g}^{0}$ the states resemble free electrons
in the field of {\em uncorrelated quantum fluctuations} of the
lattice \cite{braz-84}; here the self-trapping has not enough time
to be developed. But approaching the exact threshold $E_{g}$, the
excitations evolve towards eigenstates which are {\em
self-trapped} {\it e-ph} complexes. The PGs must be common in $1D$
semiconductors just because of favorable conditions for
self-trapping \cite{rashba}. The PG is especially pronounced when
the bare gap is opened spontaneously as a symmetry breaking
effect. In quasi-$1D$ conductors it is known as the
Peierls-Fr\"{o}hlich instability leading to the CDW formation.
Here the PGs were addressed experimentally by means of optics
\cite{optics}, PES \cite{pes} and by the new methodics of the
coherent interlayer tunneling \cite{tunneling}.

Detailed theories of the subgap absorption in optics have been developed
already for systems with low symmetries (nondegenerate, like semiconductors
with gapful phonons, or discretely degenerate like the dimerized Peierls
state): for a general type of polaronic semiconductors \cite{iosel} with an
emphasis to long range Coulomb effects and for the one dimensional Peierls
system with an emphasis to solitonic processes \cite{kiv-86}. Recently the
authors \cite{matv-01} extended the theory of pseudogaps to single
electronic spectra in application to the PES and, particularly intriguing,
to the ARPES (momentum resolved PES) probes. But properties of ICDWs are
further complicated by appearance of the gapless collective mode which bring
drastic changes. The case of acoustic polarons in 1D semiconductors belongs
to the same class.

The specifics of $1D$ systems with continuous degeneracy
(with respect to the phase for the ICDW, to displacements for usual
crystals) is that even single electronic processes can create topologically
nontrivial excitations, the solitons. Thus for the ICDW a single electron or
hole with the energy near the gap edges $\pm \Delta _{0}$ spontaneously
evolves to the nearly amplitude soliton - {\it AS }whilst the original
electron is trapped at the local level near the gap center \cite{braz-84}.
The energy $\approx 0.3\Delta _{0}$ is released, at first sight within a
time $\omega _{ph}^{-1}$ $\sim 10^{-12}s$. We will see that actually there
is also a long scale adaptation process which determines transition
probabilities. Similarly, the usual acoustic polaron in a 1D semiconductors
is characterized by the electronic density $\rho $ selflocalized within the
potential well $\sim \partial \varphi /\partial x\sim \rho $, hence a finite
increment $\varphi (+\infty )-\varphi (-\infty )\neq 0$ of the lattice
displacements $\varphi (x)$ over the length $x$ which is the signature of
topologically nontrivial solitons.

\section{Instanton approach to pseudogap}

We shall use the adiabatic approximation valid when changes of
electronic energies are much bigger than relevant phonon
frequencies. Electrons are moving in the slowly varying potential,
e.g. ${\rm Re}\{\Delta (x,t)\exp[i2k_{F}x]\}$ for the ICDW. So at
any instance $t$ their energies $E(t)$ and wave functions $\psi
(x,t)$ are defined as eigenstates for the instantaneous lattice
configuration and they depend on the time $t$ only parametrically.
The PES absorption rate $I(\Omega )$ for the light frequency
$\Omega $ can be written as the functional integral
\[
I(\Omega )\propto \int_{0}^{\infty }dT
\int D[\Delta (x,t)]\psi _{0}(0,T)\psi_{0}^{+}(0,0)\exp (-S)
\]
were $\psi _{0}$ is the wave function of the particle (for the PES it is
actually a hole) added and extracted in moments $0$ and $T$ at the
fluctuational intragap level $E_{0}$. The time $t,T$ is already chosen along
the imaginary axis where the saddle points of this integral are commonly
believed to be found (see \cite{matv-01,iosel} for references). The
Euclidean action
\[
S[\Delta (x,t),T]={(\int_{-\infty }^{0}+\int_{T}^{\infty})dt\,
L_{0}+\int_{0}^{T}dt\,(L_{1}-\Omega )}
\]
is determined by Lagrangians $L_{j}$ where the labels $j=0,1$
correspond to ground states for $2M$ (the bare number) and $2M\pm
1$ electrons in the potential $\Delta (x,t)$. For calculations of
subgap processes only the lowest singly filled localized state is
relevant which energy $E_{0}$ is split off inside the gap, hence
$L_{1}=L_{0}+E_{0}$. The main contribution comes from saddle
points of $S$, the instantons, which are extremas with respect to
both the function $\Delta(x,t)$ and the time $T$, the last
condition yielding $E_{0}(T)=\Omega$.

\section{Creation of spin solitons in Incommensurate CDWs.}
For the ICDW the order parameter is the complex field $\Delta
=|\Delta(x,t)|\exp [i\varphi (x,t)]\,$ acting upon electrons by
mixing states near the Fermi momenta points $\pm k_{F}$. The
Lagrangians $L_{j}$ consist of the bare kinetic
$\sim |\partial_{t}\Delta |^{2}$ and potential $\sim |\Delta|^{2}$ lattice
energies and of the sum over the filled electron levels:
\[
L_{j}=2/\pi v_{F}\int dx|\partial _{t}\Delta |^{2}/\omega
_{0}^{2}+V_{j}[\Delta (x,t)]\,
\]
where $v_{F}$ is the Fermi velocity in the metallic state and $\omega
_{0}\ll \Delta _{0}$ is the amplitude mode frequency in the CDW state. The
stationary state with an odd number of particles, the minimum of $V_{1}$, is
the amplitude soliton ($AS$) $\Delta \Rightarrow -\Delta $ with the midgap
state $E_{0}=0$ occupied by the singe electron. The evolution of the free
electron to the AS can be fortunately described by the known \cite{braz-84}
exact solution for intermediate configurations characterized by
the singly occupied intragap state $E_{0}=\Delta _{0}\cos \theta $ with
$0\leq \theta \leq \pi $, hence $-\Delta _{0}\leq E_{0}\leq \Delta _{0}$. It
was found to be the Chordus Soliton ({\it ChS}) with $2\theta $ being the
total chiral angle: $\Delta (+\infty )/\Delta (-\infty )=\exp (2i\theta )$.
The filling number of the intragap state $\nu =0,1$ corresponds to labels
$j=0,1$. The term $V_{0}(\theta )$ increases monotonically from $V_{0}(0)=0$
for the $2M$ ground state (GS) to $V_{0}(\pi )=2\Delta _{0}$ for the $2M+2$
state with two free holes. The term $V_{1}(\theta )=V_{1}(\pi -\theta )$ is
symmetric describing both the particle upon the $2M$ GS and the hole upon
the $2M+2$ GS. Apparently $V_{1}(0)=V_{1}(\pi )=\Delta _{0}$ while the
minimum is reached at $\theta =\pi /2$ that is for the purely AS:
$\min V(\theta )=$ $V_{1}(\pi /2)=W_{s}<\Delta _{0}$ where
$W_{s}=2/\pi \Delta _{0}$ is the AS energy.
To create a nearly AS with $\theta =90^{\circ }$, the
light with $\Omega \approx W_{s}$ is absorbed by the quantum fluctuation
with $E_{0}(\theta )=W_{s}$ which is close to the chordus soliton with the
angle $\theta \approx 50^{\circ}$.

As the topologically nontrivial object, the AS cannot be created in a pure
form: adaptational deformations must appear to compensate for the
topological charge. These deformations are developing over long space-time
scales and they can be described in terms of the gapless mode, the phase
$\varphi $, alone. Hence allowing for the time evolution of the chiral angle
$\theta \rightarrow \theta (t)$ within the core, we should also unhinder the
field $\varphi \rightarrow \varphi (x,t)$ at all $x$ and $t$. Starting from
$x\rightarrow -\infty $ and returning to $x\rightarrow \infty $ the system
follows closely the circle $|\Delta |=\Delta _{0}$ changing almost entirely
by phase. Approaching the soliton core the phase matches approximately the
angles $\pm \theta $ which delimit the chordus part of the trajectory. The
whole trajectory is closed which allows for the finite action $S$. The
chordus angle $2\theta (t)$ evolves in time from $\theta (\pm \infty )=0$ to
$\theta _{m}$ in the middle of the $T$ interval. For $T\rightarrow \infty $,
that is near the stationary state of the AS, $\theta _{m}\rightarrow \pi /2$.
Actually this value is preserved during most of the $T$ interval so that
changes between $\theta =0$ and $\theta =\pi /2$ are concentrated within
finite ranges $\tau \sim \xi _{0}/u\ll T$ near the termination points. From
large scales we view only a jump
$\varphi (x,t)\approx \theta (t){\rm sgn}(x)$ with
$\theta (t)\approx \pi /2\Theta (t)\Theta (T-t)$ for a well developed
AS. Since the configuration stays close to the AS during the time $T$, the
main core contribution to the action is
\[
S_{core}=(W_{s}-\Omega )T+\delta S_{core}
\] where the first correction
$\delta S_{core}^{0}=cnst$ comes form
regions around moments $0,T$ independently. The significant $T$ dependent
contribution $\delta S(T)$ comes from interference of regions $0$ and $T$
which for the ICDW communicate via the gapless phase mode. Its effect can be
easily extracted if we generalize the scheme suggested earlier for static
solitons at presence of interchain coupling \cite{soliton3d}. The action for
the phase mode is
\begin{equation}
S_{snd}[\varphi (x,t),\theta (t)]=
\frac{v_{F}}{4\pi }\int \int dxdt\left((\partial _{t}\varphi /u)^{2}
+(\partial _{x}\varphi )^{2}\right)
\label{s-snd}
\end{equation}
where $u$ is the phase velocity. The chordus soliton forming around $x_{s}=0$
enforces the discontinuity $\varphi (t,\pm 0)=\mp \theta (t)$. Integrating
out $\varphi (x,t)$ from $\exp \{-S_{snd}[\varphi ,\theta ]\}$ we arrive for
$\theta (t)$ at the typical action for the problem of quantum dissipation
\cite{leggett}
\[
S_{snd}\{\theta ]\approx -\frac{v_{F}/u}{2\pi ^{2}}\int \int dt_{1,2}
\dot{\theta}(t_{1})\ln |(t_{1}-t_{2})|\dot{\theta}(t_{2})
\]
that is $S\sim \sum |\omega ||\theta _{\omega }|^{2}$. The dissipation comes
from emission of phase phonons while forming the long range tail in the
course of the chordus soliton development. This action, together with $V_j$,
can be used to prove the above statements on the time evolution of the ChS core.

Remember now that $\partial_{t}\theta $ is peaked within narrow regions $\sim \xi _{0}/u$
around moments $t=0$ and $t=T$ and close to zero otherwise. Then
\begin{equation}
S_{snd}\approx \left( v_{F}/4u\right) \ln (uT/\xi _{0})  \label{log}
\end{equation}
The picture is clear in the space-time domain. The AS creates the $\pi -$
discontinuity along its world line: $(0<t<T,x=0)$. To be topologically
allowed, that is to have a finite action, the line must terminate with half
integer vortices located at $(0,0)$ and $(0,T)$ which circulation provides
the jump $\delta \varphi =\pi $ along the interval compensating for the sign
change of the amplitude ($\Delta \Rightarrow -\Delta $ combined with
$\varphi \Rightarrow \varphi +\pi $ leaves the order parameter
$\Delta \exp(i\varphi )$ invariant, see \cite{comb-top} for discussion of ''combined
topological defects''). Then the standard energy of vortices for (\ref{s-snd})
leads identically to the action (\ref{log} ). Contrary to usual $2\pi -$ vortices,
the connecting line is the physical singularity which tension gives $S_{core}$.

Minimizing $S_{tot}=S_{core}+S_{snd}\,$with respect to $T$, we obtain near
the AS edge $\Omega \geq W_{s}$ the power law
\begin{equation}
I(\Omega )\propto \left( \frac{\Omega -W_{s}}{W_{s}}\right) ^{\beta}
\,,\;\beta =\frac{v_{F}}{4u}  \label{beta}
\end{equation}
which is much more pronounced than the nearly exponential law \cite{matv-01}
for gapful cases: $I\sim \exp \{-\delta \Omega \ln (\delta \Omega )\}$.

Behavior{\bf \ near the free edge }$\Omega \approx \Delta _{0}$ is
dominated by small fluctuations $\eta $ of the gap amplitude
$|\Delta |=\Delta_{0}+\eta $ and of the Fermi level
$\delta E_{F}=\varphi ^{\prime }v_{F}/2$ via the phase gradient $\varphi
^{\prime }=\partial _{x}\varphi $. We shall analyze it in a frame
of a generic problem of the combined (gapful and acoustic) polaron
where the simpler single particle formulation allows for a more
detailed analysis.

\section{$1D$ semiconductors: from polarons to quantum noise.}

Consider electron (hole) states in a 1D dielectric near the edge of a
conducting (valence) band. We shall take into account the gapful phonon mode
$\eta $ with the coupling $g_{0}$ and the sound mode (for which we shall
keep the ''phase'' notation $\varphi (x,t)$) with the velocity $u$ and the
coupling $g_{s}$. In generic semiconductors the sound mode is always present
as the acoustic phonon while the gapful one can be present as an additional
degree of freedom. In CDWs the gapful mode is always present as the
amplitude fluctuation $|\Delta |=\Delta _{0}+\eta $ while the sound mode
appears in the ICDWs as the phase $\Delta =|\Delta |\exp [i\varphi ]$.

Within the adiabatic approximation the action $S$ has the form
\begin{eqnarray}
&&\int dxdt\left[ \left( \frac{1}{2m}\left| \Psi ^{\prime
}\right| ^{2}-\Omega \left| \Psi \right| ^{2}\right) +\left( g_{s}\varphi
^{\prime }+g_{0}\eta \right) \left| \Psi \right| ^{2}\right] +  \nonumber \\
&&\int dxdt\left[ \frac{K_{s}}{2}
\left( \dot{\varphi}^{2}/u^{2}+\varphi ^{2}\right) +
\frac{K_{0}}{2}\left( \dot{\eta}^{2}/\omega_{0}^{2}+
\eta ^{2}\right) \right] \,.  \label{S-pol}
\end{eqnarray}
(For the PES problem, the electron wave function $\psi=psi(x,t)$ is not zero only at $0<t<T$.
 Thus for the ICDW case we have $m=\Delta _{0}/v_{F}^{2}$, $g_{0}=1$,
$g_{s}=v_{F}/2$, $K_{s}=v_{F}/2\pi $, $K_{0}=4/\pi v_{F}$,
$2^{3/2}u/v_{F}=\omega _{0}/\Delta _{0}$ and $\Omega $ is counted with
respect to $\Delta _{0}$. {\em \ }The stationary state, the time independent
extremum of (\ref{S-pol}), corresponds to the selftrapped complex, the
polaron \cite{rashba}. Here it is composed equally by both $\eta $ and
$\varphi ^{\prime }$ which contribute additively to the
coupling $\lambda =\lambda _{s}+\lambda _{0}=g_{s}^{2}/K_{s}+g_{0}^{2}/K_{0}$.
The polaronic length scale $l$ for $\eta \sim \varphi ^{\prime }\sim |\Psi
|^{2}\equiv \rho _{p}(x)$ is $l=2/m\lambda $ and the total energy is
$W_{p}=-m\lambda ^{2}/24$ . The conditions $|W_{p}|\gg \omega _{0}$ and
$\lambda \gg u$ define the adiabatic approximation. For
the CDW case $\lambda _{s}=v_{F}\pi /2$ and $\lambda _{0}=v_{F}\pi /4$ hence
$\lambda \sim v_{F}$ and we would arrive at $|W_{p}|\sim \Delta _{0}$ and
$l\sim \xi _{0}=v_{F}/\Delta _{0}$ which are the microscopic scales where the
single electronic model may be used only qualitatively. The full scale
approach for nearly stationary states has been considered above, but the
upper PG region near the free edge $\Delta _{0}$ will be described by the
model (\ref{S-pol} ) even quantitatively.

We can integrate out the fields $\varphi $ and $\eta $ at all $x$,$t$ to
obtain the action $S\{\Psi ;T\}$ in terms of $\Psi $ alone which is defined
now only at the interval $(0,T)$:

\begin{eqnarray}
&&\int dxdt\left( \frac{1}{2m}\left| \partial _{x}\Psi \right| ^{2}-\Omega
\left| \Psi \right| ^{2}\right) -
\frac{1}{2}\int \int dt_{1,2}\int \int dx_{1,2}  \label{S-sT} \\
&&\left\{ U_{s}(\delta _{1,2})\rho ^{\prime }(x_{1},
t_{1})\rho ^{\prime}(x_{2},t_{2})\right\} +U_{0}(\delta _{1,2})
\rho (x_{1},t_{1})\rho(x_{2},t_{2})  \nonumber
\end{eqnarray}
where $\delta _{1,2}=(x_{1}-x_{2},t_{1}-t_{2})$ and
\begin{equation}
U_{s}=\frac{\lambda _{s}u}{2\pi }\ln \sqrt{x^{2}+t^{2}u^{2}}\,,\;U_{0}
=\frac{1}{2}\lambda _{0}\omega _{0}\exp [-\omega _{0}|t|]\delta (x)  \label{Us-U0}
\end{equation}

An equivalent form for $S\{\Psi ;T\}$, suitable at large $T$, is obtained
via integrating by parts:
\begin{equation}
\int dx\int dt\left( \frac{1}{2m}\left| \Psi ^{\prime }\right| ^{2}-\Omega \rho
-\frac{\lambda }{2}\rho ^{2}\right) +
\frac{1}{2}\int dt_{1,2}\int dx_{1,2}\dot{\rho}(1)U(\delta _{1,2})\dot{\rho}(2)
\label{S-lT}
\end{equation}
where $U(x,t)=u^{-2}U_{s}+\omega _{0}^{-2}U_{0}$.

{\bf The absorption near the absolute edge} $\Omega \approx W_{p}$ is
determined by the long time processes when the lattice configuration is
almost statically self-consistent with electrons. The first term in
(\ref{S-lT}) is nothing but the action $S_{st}$ of the static polaron which
extremum at given $T$ is
$S_{st}\approx -T\delta \Omega \,,\;\delta \Omega=\Omega -W_{p}$.
The second term in (\ref{S-lT}) $S_{tr}$ collects
contributions only from short transient processes near $t=0,T$ which are
seen as
$\partial _{t}\rho (x,t)\approx \rho _{p}(x)[\delta (t)-\delta (t-T)]$
where $\rho _{p}$ is the density for the static polaron solution. We obtain
\begin{eqnarray*}
S_{tr} &\approx &\int \int dx_{1,2}\rho _{p}(x_{1})
\rho_{p}(x_{2})U(x_{1}-x_{2},T) \\
&=&\frac{\lambda _{s}}{2\pi u}\ln \frac{uT}{l}+C_{0}
\frac{\lambda _{0}/l}{\omega _{0}}\exp [-\omega _{0}T]+cnst
\end{eqnarray*}
with $C_{0}\sim 1$. We see the dominant contribution of the sound mode which
grows logarithmically in $T$ while the part of the gapful mode decays
exponentially. If the sound mode is present at all, then the extremum over
$T$ is
\[
T\approx \frac{\lambda _{s}}{2\pi u}\frac{1}{\delta \Omega }
\,,\;S\approx\frac{\lambda _{s}}{2\pi u}\ln
\frac{C_{s}|W_{p}|}{\delta \Omega }\,,\;C_{s}\approx 0.9
\]
We find that near the absolute edge $\Omega \approx W_{p}$ the
absorption is dominated by the power law $I\sim |\delta \Omega
/W_{p}|^{\alpha }$, with a big index $\alpha =\lambda _{s}/(2\pi
u)$. For parameters of the ICDW we obtain $\alpha
=v_{F}/4u=\equiv\beta $ in full accordance with the exact
treatment (\ref{beta} ).

Only in absence of sound modes $\lambda _{s}=0$ the gapful contribution can
determine the absolute edge. Then the minimization over $T$ of
$S=S_{core}+\delta S_{gap}$ would lead qualitatively to the result of
\cite{matv-01}:
\[
I\sim \exp \left( cnst\frac{\Omega -W_{p}}{\omega _{0}}\ln
\frac{W_{s}}{(\Omega - W_{s})} \right)
\]
for the intensity near the absolute edge.

{\bf The opposite regime near the free edge }$\Omega \approx 0$
($\Omega\Rightarrow \Omega -\Delta _{0}$ for the ICDW) is dominated by fast
processes of quantum fluctuations. Their characteristic time $T=T(\Omega )$
is short in compare to the relevant phonon frequency: $T\ll \omega _{0},u/L$,
where $L=L(\Omega )$ is the characteristic localization length for the
fluctuational electronic level at $E_{0}=\Omega $. Neglecting time
variations within the short interval $(0,T)$, we can estimate the action
(\ref{S-sT}), term by term, as
\[
S\approx \frac{C_{1}T}{mL^{2}}-\Omega T-C_{2}\lambda _{s}u
\left( \frac{T}{L}\right) ^{2}-C_{3}\lambda _{0}\omega _{0}\frac{T^{2}}{L}
\]
with $C_{i}\sim 1$. Its extremum over both $L$ and $T$ yields
\begin{equation}
S\sim \frac{|\Omega |^{3/2}/m^{1/2}}{\max
\left\{ |m\Omega |^{1/2}u\lambda_{s};\omega _{0}\lambda _{0}\right\}}
\label{both}
\end{equation}
Then $I\sim \exp(-S)$ interpolates between the closest
$S\sim |\Omega |^{3/2}$
and the more distant
$ S\sim |\Omega |$ vicinities of
the free electronic edge. For the purely acoustic case $\lambda _{0}=0$ a
variational estimation for the numerical coefficient as
$C_{1}\approx 1/6,C_{2}\approx 0.06$ gives

\begin{equation}
I\sim \exp [-cnst|\Omega |/mu\lambda _{s}]\,,\;cnst\approx 2.8  \label{exp}
\end{equation}

\section{Conclusions.}

 We conclude that the subgap absorption in systems with
gapless phonons is dominated by formation of long space-time tails
of relaxation. It concerns both the acoustic polarons in $1D$
semiconductors and solitons in ICDWs. Near the free edge the
simple exponential, Urbach type, law (\ref{exp}) appears competing
with stretched exponential ones typical for tails from optimal
fluctuations, see (\ref{both}). The deeper part of the PG is
dominated by the power law (\ref{beta}) (with the big dependable
index) singularity near the absolute edge.

Our results have been derived for single electronic transitions: PES and tunneling.
They can also be applied to intergap (particle-hole) optical transitions
as long as semiconductors are concerned.
For the ICDW results are applied to the free edge vicinity.
But the edge at $2E_s$ will disappear in favor of optically active gapless phase mode.

\acknowledgements {\it S. M. acknowledges the hospitality of the
Laboratoire de Physique Theorique et des Modeles Statistiques
(Orsay, France) and the support of the CNRS and the ENS - Landau
foundation. This work was partly performed within the INTAS grant
\#2212.}

\end{document}